# Investigating Girls' Perspectives and Knowledge Gaps on Ethics and Fairness in Artificial Intelligence in a Lightweight Workshop

Jaemarie Solyst, Alexis Axon, Angela B. Stewart, Motahhare Eslami, and Amy Ogan
jsolyst@andrew.cmu.edu, aaxon@andrew.cmu.edu, angelast@andrew.cmu.edu,
meslami@andrew.cmu.edu, aeo@andrew.cmu.edu
Carnegie Mellon University

**Abstract:** Artificial intelligence (AI) is everywhere, with many children having increased exposure to AI technologies in daily life. We aimed to understand middle school girls' (a group often excluded group in tech) perceptions and knowledge gaps about AI. We created and explored the feasibility of a lightweight (less than 3 hours) educational workshop in which learners considered challenges in their lives and communities and critically considered how existing and future AI could have an impact. After the workshop, learners had nuanced perceptions of AI, understanding AI can both help and harm. We discuss design implications for creating educational experiences in AI and fairness that embolden learners.

## Introduction

Youth have regular direct exposure to and face the consequences of Artificial Intelligence (AI), including through their use of social media and smart home technologies. However, unfairness and bias in AI that affects specific users has been well-documented (e.g., O'Neil, 2016, Buolamwini, et al., 2018), including digital assistants' difficulty comprehending or responding to younger users (e.g., Druga et al., 2017) and users with accents (Lima et al., 2019). AI also introduces concerns about families' privacy around data collection (e.g., Lau et al., 2018). Children's frequent interfacing with AI-based technology, and therefore its potential harms, calls for the creation of more learning opportunities for youth to become AI literate and aware of its risks and challenges. This becomes more essential for underserved groups in tech who are more likely to be harmed by biased algorithms. Girls are often a population excluded from tech design and education (Sultan et al., 2018) yet can be greatly impacted by algorithmic bias (e.g., Buolamwini, et al., 2018). It is important to focus on underserved groups in AI education, so they avoid misinformation, have learning resources, and lend insight into a more just future.

Researchers and curriculum developers have recently begun to create learning opportunities to address children's knowledge gaps in AI literacy. However, many focus primarily on the technical aspects, excluding analysis of the ethical implications of this technology. Additionally, learners in this work have often been predominantly male (e.g., Williams et al., 2019; Kahn et al., 2018), leaving researchers and those who develop educational opportunities with a limited understanding of how girls perceive and learn about AI, potentially compounding existing inequities in AI representation. Some research has looked at teaching K-12 AI with a focus on ethics and fairness, including interpreting the agendas of various stakeholders in AI technology design (Payne, 2019; DiPaola et al., 2020; Register & Ko, 2020) and family-based education around exploring, manipulating, and analyzing AI (Chklovski et al., 2019; Druga et al., 2021). Touretzsky et al. (2019) and Long & Magerko (2020) describe ideas that children should learn about AI, including technical aspects and ethics surrounding impacts on humans and society. Ali et al. (2021) explored teaching middle schoolers about generative AI and corresponding ethical considerations, finding it was necessary to take technical, applied, and ethical approaches. We build on prior work by exploring underserved learners' perceptions, opinions, and knowledge gaps around AI technology to leverage in creating learner-centered education, as well as explore a shorter workshop format.

In this work, we focused on middle school girls, with middle school being a prime time to develop STEM identity (Sadler et al., 2012). To explore middle school girls' perceptions, knowledge gaps, and opinions to inform future creation of AI literacy materials, we created a lightweight workshop with content that assumed no prior knowledge in computing and AI, encouraged critical and creative thinking about AI systems and ethics, and supported learners in thinking about training data. The lightweight (under 3 hours) workshop, compared to other programs that are longer and more intensive, was aimed at inclusivity, reducing common barriers to out-of-school time computing participation like time commitment, commute, etc. We aim to expand upon prior work by integrating both technical and ethical aspects of AI, first taking inspiration from Touretzky et al. (2019) and Long & Magerko (2020), in particular, learning from data and nuanced societal impacts. We took an asset- rather than a deficit-based approach (Garoutte & McCarthy-Gilmore, 2014) which involves centering learners' prior knowledge and interests as a base for further learning and engagement (Coleman & Davis, 2020). This approach is particularly appropriate to successfully support girls of diverse backgrounds in learning computing (Scott et al., 2015). To conduct our research, we used a lens of futuring in our activities, which is a method that encourages reflection and analysis around technology and problem-solving by speculating about possible future tools (Dunne & Raby, 2013). Through futuring, we also aimed to gain insight into what is motivating and important to learners, building on their existing knowledge. We strive to address the following research questions: (1) What are middle school girls' conceptions of AI and ethical considerations? (2) How do learners apply their knowledge to ideate and critically analyze AI technologies in short workshops? To create effective and empowering educational opportunities for these learners to gain information about both the technical and ethical aspects of AI, it is essential to understand and center their current funds of knowledge. We report findings about middle school girls' understandings of AI and ethics before, during, and after the lightweight workshop, then offer design implications.

## Methods

### Data collection and participants

This study took place over five months as a standalone workshop module situated within five individual computing camps (labeled Groups A-E), each held over three to five days. We took a design-based, iterative approach between each camp, slightly adjusting or adding activities to improve the workshop. Each program focused on elementary programming within the context of robotics and emphasized power and identity in techmaking. The programs and workshop sessions did not assume any prior experience with computing or AI. After Group C, we added an additional open-ended final project in a 90-minute session held two days after the initial AI and fairness session. We collected ~14 hours of video and interview data overall with the 5 groups.

*Participants*: Middle school girls ($N = 32$, see Table 1) were recruited through a robotics-focused organization, "BoltGirls," in the east coast of the US and a general girls organization, "AmazingGirls," in the southwest of the US. Girls recruited from BoltGirls had more prior experience with computing than girls recruited from AmazingGirls. Each camp included participants from either BoltGirls or AmazingGirls.

*Data collection*: The AI workshop with educational content and activities lasted about an hour and a half in duration for all groups. In our first group (A) we started with one AI-related session, after which we divided the session into two, but due to the drop-in nature of informal learning experiences, some learners only participated in one of the two sessions. Sessions were conducted synchronously online over Zoom (https://zoom.us) and recorded with chat logs saved. We utilized Jamboards for more efficient group work with peers and a common environment for learners and instructors to observe what others were doing during the virtual sessions, and the content of these boards was collected and analyzed. We also collected data on learners' prior knowledge in all groups via warm-up activities; to facilitate data collection, we decided to distribute these questions as a short pre-survey for groups D and E ($N = 17$), which asked them to give their best definition of AI, give two examples of AI, and their opinion on whether AI is smarter than them. This pre-survey was added for groups D and E as part of our iterative processes, and 3 of the 17 learners opted to answer a subset of the questions. We also invited learners from groups D and E to participate in follow-up semi-structured interviews, and 10 learners agreed. In interviews, we again asked for their best definition of AI and whether AI is smarter than them, and also what fair and unfair technology is, and where they may have gotten inspiration about AI (e.g., in the media or news). In all questions in the pre-survey and follow-up interviews, we also asked about their reasoning for their answers.

### Qualitative coding

We conducted thematic analysis using inductive open coding (Corbin & Strauss, 1990) to uncover themes in video recordings, chat logs, survey and interview data, and learning artifacts (Jamboards and slide decks). Two researchers independently developed a list of themes using affinity diagramming (Beyer & Holtzblatt, 1999), with each source of data contributing to a single set of themes. They then met to discuss, combine, and justify themes, before discussing them with the other authors. The thematic analysis was iterative across camps, as additional data was collected. The researchers discussed the themes regularly with each other and the rest of the research team, until the team agreed on the final set of findings. The findings below describe major themes from our study, which address our research questions. The supporting data for each theme is derived from prompts in the learning activities, unless otherwise specified that the insights are from the pre-survey or follow-up interviews.

### Learning content and activities

The learning goals for our workshop sessions included understanding AI and supervised machine learning on a high level and recognizing that AI can have nuanced (i.e., both helpful and harmful) societal impacts. Learners started the session by discussing what they thought the word 'bias' meant in a group discussion. We then played a guessing game, where learners guessed if the technology pictured has AI (e.g., personalized recommendations) and what technology does not (e.g., a hairdryer) and explained their reasoning. To understand AI, learners need a working definition of an algorithm, which we gave in a short lesson covering how algorithms consist of three parts: an input, steps to change the input, and an output (adapted from Payne, 2019). After iterating our materials, we created an application activity for groups D and E to practice in an activity where they filled out sticky notes for their favorite food (output), with ingredients (input) and preparation or cooking actions (steps to change input).

**Table 1**
*Participant demographics.*

| Group | N | Had prior exp | Organization | Ages | Reported race |
|---|---|---|---|---|---|
| A | 4 | 2 | BoltGirls | 12-14 | Asian (1), Black (1), white (1), Asian & white (1) |
| B | 3 | 1 | AmazingGirls | 12-14 | Hispanic/Latina(1), white (1), Hispanic/Latin & white (1) |
| C | 8 | 6 | BoltGirls | 11-13 | Asian (5), white (3) |
| D | 4 | 2 | AmazingGirls | 11-13 | Latina/Hispanic (1), white(2), did not say (1) |
| E | 13 | 10 | BoltGirls | 12-14 | Asian (7), Black (1), Indian American/First Nation (1), white (2), did not say (2) |

We then shifted over to AI, providing a definition of what AI is and how it relates to algorithms. We centered this lesson on supervised machine learning, emphasizing the concept of training data, since it is a pressing and relevant topic related to power and unfairness (O'Neil, 2016), as well as an important topic to cover in AI education (Touretzsky et al., 2019; Long & Magerko, 2020). Learners then played Google's QuickDraw (quickdraw.withgoogle.com), an interactive, short game leveraged in other AI education programs, e.g., Druga et al., 2021. We opted for this over other tools (e.g., Google's Teachable Machine) to fit within our lightweight workshop. We then discussed why and how the algorithm identifies or misidentifies their drawings, sifting through the data set that other users drew. While most groups concluded that the algorithm may not be able to identify certain drawings because they are not represented in the training data, the instructor made sure to emphasize this point before moving on. The learners then watched a 4-minute video on algorithmic bias stemming from systemic bias in society and data, discussing examples where biased data in AI could do great harm. Learners were then prompted to consider where else harmful effects of algorithmic bias could take place. Learners then took part in two Jamboard group activities and an individual, guided final project, which we go into more detail next.

After the lesson and discussions on AI and bias, we ran several activities designed to encourage learners to critically analyze technologies and apply concepts delivered in the content. We use artifacts generated from these activities as one of our main sources of data, alongside pre-surveys, follow-up interviews, video recordings, and chat logs. In Jamboard activities, the girls contributed to at least one sticky note and discussed the prompt. The facilitator helped with pacing and prompting. If girls did not want to write, they also had the option to type in the Zoom chat or speak, and the facilitator would fill out a sticky note for them.

In Activity 1, *Digital Assistants Help and Harm*, we first asked learners to consider how and who existing digital assistant technology, such as Alexa, Siri, or Google Home, could help and harm. We started by introducing what digital assistants are and brainstorming together what they are capable of. We then provided a Jamboard with prompts. Instructors helped to pace the learners through the Jamboard and offered an example when the learners were stuck, such as accent bias in voice recognition, but aimed not to scaffold too heavily. Jamboard prompts included inquiring what fair and unfair interactions digital assistants could do and who would be affected, as well as which training data to include or exclude for more equitable interactions (see Figure 1). In Activity 2, *Community AI Robot,* learners then considered a challenge in their community that they wanted to solve and were prompted to brainstorm a robot AI that could be a part of the solution. We chose a robot here as part of the prompt due to the context of the computing camps, since robotics were an emphasized theme. Similar to Activity 1, learners then brainstormed who it could benefit or be harmed by their AI robot. They then ideated on which training data they would include or exclude to facilitate fair interactions or avoid unfair interactions.

For groups D and E, we added a final project, which took place two days after the initial workshop session and involved learners brainstorming any futuristic AI technologies to help solve a problem that they are passionate about. On a Jamboard, we started by brainstorming AI technologies and what they can do (e.g., types of personalization, face/screen/object recognition, etc.). Then, learners brainstormed together different challenges in the world, and finally considered how the AI examples they came up with could be applied to support solving some of these challenges. After the group-wide brainstorm, they were given their own slide deck to create a presentation on a futuristic AI technology of their choice. Slide prompts asked learners to define the problem their AI technology could help solve and describe how and who the AI technology could help, what form the AI technology would take (e.g., a robot, website, app, etc.), how the AI could do harm and to whom. They then were further prompted to identify fair and unfair interactions that they wanted their AI technology to have and avoid, as well as what training data they would include or exclude to facilitate or mitigate such interactions (see Figure 1). Learners had opportunities to present their progress throughout the session.

**Figure 1**
*Two screenshots of learners' final projects analyzed in this study. Left: learner suggests a human-in-the-loop solution. Right: a learner suggests that AI may perpetuate inequity by being too expensive.*

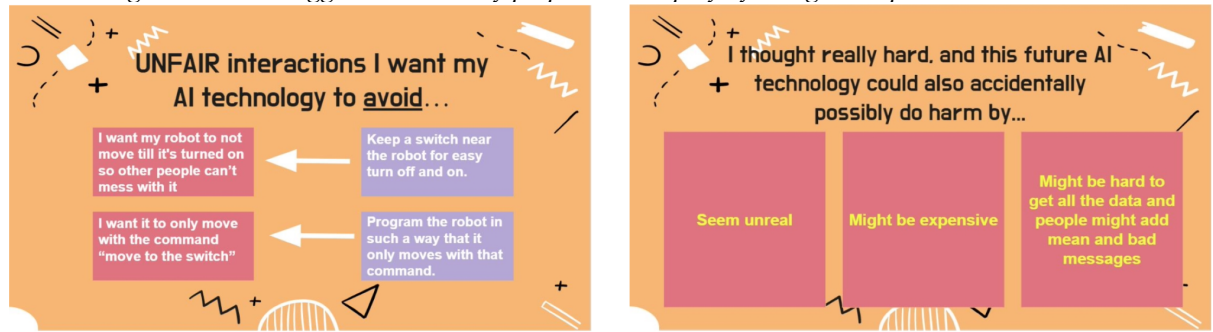

## Findings

### (Mis)perceptions of AI: Understanding of AI

We wanted to understand learners' conceptions of AI before engaging in the workshop and consider how they shifted over the course of the experience. We consider a working definition of AI to be along the lines of e.g., 'algorithms that can learn to make meaningful output or decisions.' We observed that almost all 17 learners who took the pre-survey were able to give an example of a technology that uses AI, with 11 giving two. For instance, learners listed digital assistants (i.e., Siri, Google Home, ($n$ = 7), self-driving cars ($n$ = 2), and facial recognition ($n$ = 2) as examples of AI. Some learners ($n$ = 3) were able to give more technical definitions of AI, such as "*something that is developed by humans, to make robots think and solve problems.*" or "*when a form of technology is able to do something without another person programming it to do that.*" Many ($n$ = 12) gave definitions that were not technical but showed some awareness of the field such as relevance to programming, brains, or robots. Several learners indicated uncertainty in their responses (e.g., responding with a question mark). Afterward, some learners showed a more complex understanding of AI in the follow-up interviews, with 4 (of 7) learners able to give a technical definition of AI, of which two had already given working definitions initially. While the workshop may not have supported all learners in clearly defining AI, most learners in the follow-up interviews drew on new concepts and showed an increased ability to talk about AI and related topics.

      We were further interested in understanding learners' mental models of artificial intelligence. For instance, some misconceptions potentially arise from believing machines have no faults, with past work finding that by age 7-10, children believe that AI is "smarter than they are" (Druga et al., 2018). Our middle school-age learners had varied perceptions to this question, e.g., in the second half of the pre-survey when learners were asked if AI was smarter than them, most (10 of 14) answered that `yes,' they thought so. However, by the time of the follow-up interviews, half (5 of 10) responded 'no' or 'neither' was smarter than the other. For those who believed AI was smarter, their reasoning related to intelligence comprising informational knowledge and facts, a finding consistent with Druga et al., 2017. On the other hand, the learners who considered themselves "smarter" than AI mentioned reasons such as the fact that humans build AI, and recognized that this creates limitations in AI capabilities, e.g., "*if the AI wasn't very trained at something, then a human would probably find a more accurate solution*". Other girls held a more expansive view of intelligence, that AI "*doesn't necessarily have emotions or think the exact way that I would think...*" Although some of their ideas about AI may not be completely unpacked, we found that all learners, even those with low prior knowledge in AI, were able to engage with and hypothesize about AI. Specifically, after the workshop, learners displayed more nuanced views of the concept of intelligence, and compared AI to human intelligence, where AI may be smarter regarding certain types of knowledge (e.g., semantic) and humans smarter in terms of other knowledge and judgments (e.g., emotional intelligence). Our brief AI and algorithmic bias content provided a foundation for them to gain some new mental models and make inferences about AI and be emboldened to share ideas.

### Ethical considerations: Understanding of bias and harm

Foundational to AI and ethics is bias that arises from systemic inequity in data, as well as in the algorithms and from developers. We aimed to understand how learners understood bias. Their answers at the beginning of sessions on defining bias suggested that they generally thought that bias meant a preference for something (e.g., "*being in favor of one side or option over the other*"), which could have harmless or sometimes harmful consequences, given the context and outcome (e.g., "*it can be unfair when someone is biased without proper* *reasoning*"). This is different from prior work that suggests younger children may not know what the term bias means (Druga et al., 2021). In all groups, multiple learners were able to articulate what bias means, and the instructor further solidified it before tying it into algorithmic bias later in the workshop. Learners engaged with a number of large-scale examples of algorithmic bias in the workshop content, emphasized through discussion prompts and were able to engage in critical discussion. After watching the short video further showing examples of algorithmic bias, learners voiced that racism in AI is salient and concerning. To support critical analysis, we encouraged learners to brainstorm the benefits and harms of AI technology. They quickly identified harms that might be caused by AI, though initial ideas were not always tied into data-driven algorithmic bias.

In another take on bias, and sometimes touching on algorithmic bias, learners also identified that AI could provide misinformation (whether intentionally or not), which the learners specified could cause bias in human consumers. One learner voiced that a fair AI "*gives a neutral opinion; it's not really siding with anything.*" Learners further connected that a preference in the algorithm may affect human preference, and some suggested that AI technology give users a message as a caveat that the suggestions are only just that. Data was also a topic that is intertwined heavily with AI, which we included in our curriculum and activities. Learners were eager to discuss and bring up data, especially as it related to privacy. Multiple learners were concerned about data and how it could be collected and used (similar to Skinner et al. 2020), although they did not always tie it back to AI. For instance, when discussing existing digital assistants and chatbot technology in the Digital Assistants Help and Harm activity, learners described how data could be taken without consent or people could be hacked (an activity which could be done by a human perpetrator without AI). They also noted how large companies could profit from data, although the AI implications were not specified (e.g., using data to personalize ads), and a few learners felt that it was important systems ask permission to collect data. Lastly, we saw that most learners were able to identify stakeholders and motivations (e.g., who would create, benefit, be harmed, by tech etc.) in our Jamboard and final project activities. Some learners mentioned businesses as stakeholders and harmful technology being lucrative, suggesting that they are aware of possible harmful impacts of big companies. Thus, there are multiple avenues to encourage learners to consider stakeholders and relate these ideas to how AI can both help and harm.

## Attribution of harm and sources of awareness

To create learning experiences that encourage agency, it is important to understand how the girls perceive who has responsibility for harm caused by AI. Overall, learners had little focus on how human techmakers could be responsible for unfairness in AI beyond stealing data or intentionally providing misinformation, but rather, they often attributed harm and unfairness to the technology itself. This was particularly the case when learners mentioned embodied AI (robots) for intentional (e.g., a robot becomes 'evil,' a word learners used to describe unfair technology, and purposely does something bad) and unanticipated harm (e.g., a robot mistake resulting in injury to a human). We saw that harms learners brought up differed for existing technology versus their own future technology ideas—when the learners were asked to analyze existing digital assistant and chatbot technology in the Digital Assistants Help and Harm activity, the types of harms they identified were primarily intentional and characterized by purposely setting out to cause hurt (e.g., stealing data, spreading misinformation about a competitor, impersonating others, writing mean things online). While speculating about future robot AI they would like to build in their community, on the other hand, the harms identified were primarily unanticipated— specifically, a mistake or incorrect behavior from the AI (e.g., miscategorizing objects) rather than a bias.

Because learners brainstormed AI that was in the form of robots or had other ways of socially interfacing with humans, we observed learners' notions of fair interaction included kindness and kind language, in line with findings from Skinner et al., 2020. 'Kindness' associated with fairness and 'evil' associated with unfairness suggests that the middle school-aged participants in our study have a 'fairytale' mental model that equates fair with good and unfair with evil interactions, which may limit the ways in which learners engage with conceptualizing harmful AI. However, future work is needed to investigate how to best shift mental models to support learners in considering unanticipated (data-driven) ways AI could harm, as well as the prevalence of unanticipated bias in all technologies. Additionally, an indirect harm that learners often brought up was how AI could cause inequity indirectly by being too expensive and inaccessible to groups of people with lower income.

Sources of awareness from where learners get information about AI can inform how to relate learning content to their prior knowledge. We saw that learners' perceptions were in line with concerns presented in the media or by adults. Regular topics that came up across groups were concerns around data and privacy, and the rise of AI-driven automation taking away jobs from people, recent hot topics in the news and in line with parent perspectives (Chklovski et al., 2019). It often came up that learners believed AI could eventually become evil and turn against humans, a common sci-fi plot. All of these media-driven ideas are in line with findings in (Ali et al., 2019). However, in follow-up interviews when we asked about if any media sources influenced their ideas about AI, only 2 of 9 said that media had an impact (1 interviewee was not asked due to time constraints). This suggests that perhaps learners may not be aware of the impact of media on their perceptions of AI.

## Application of knowledge to ideate and evaluate AI: Unique and relevant ideas

Aiming to position learners as technosocial change agents and tech designers using futuring (Dunne & Raby, 2013), we investigated how learners thought of AI applications to come and found that they easily thought of pressing issues in their daily lives that were ripe for change and brainstormed future AI systems that could have a role in the solution. Despite concerns about AI noted above, learners were also very optimistic about how AI could be a part of the future. All learners who were present in all group and individual final project activities were able to consider how AI could help solve challenges. For example, learners saw possible benefits of AI in supporting accessibility needs (e.g., AI that could carry out tasks that may be difficult for a person who is physically disabled). Three specific common types of issues that learners felt passionate about in the group-wide Jamboard activities and final projects related to a) the environment and climate change (e.g., planting trees, reducing litter), b) animal well-being (e.g., rehoming animals in shelters), and c) challenges related to being a child in middle school (e.g., fitting in with peers). Many of the challenges they surfaced were highly specific to their lives (e.g., one learner focused on preventing human-environment conflict by deterring deer from eating plants in her backyard, and another on detecting crime in her neighborhood), suggesting that our asset-based approach successfully centered and worked to build off the life experiences of the learners. Creation of other AI literacy materials for middle school girls may also utilize these topics, since they may resonate with the learners.

Our data also showed that there were fewer mentions of discrimination, racism, and other forms of well-recognized unfairness that surface in algorithmic bias as issues that learners wanted to focus on in their future AI systems, despite the overarching computing program and AI curriculum mentioning some of these biases as examples (e.g., visual detection not working on all skin tones). There was less of a focus on people, identity, and nuanced types of bias overall, since in most projects and activity topics, learners chose to focus on topics related to nature or common concerns about being a middle schooler. Their brainstormed solutions also primarily focused on non-human solutions to the problem, such as a robot addressing environmental degradation by cleaning up local parks (as opposed to confronting or educating humans who litter in the first place). Across groups and activities, only one learner mentioned `sexism' as a broad problem that she was interested in using AI to solve, even though considering identity and power was a core focus of the all-girls program that the AI curriculum was a part of, and at least one salient example of sexism in AI was discussed by the instructors. Prior work has noted the contradiction that modern girls feel about sexism in their daily lives. With girls being more frequently told that they can be and do anything they want, Pomerantz et al. (2013) hypothesize that the current notion of gender equality in the US may cause young women to disregard or disbelieve existing gender injustice and threats to women's rights. Therefore, our learners may have chosen not to engage with these topics, as they may not have seen them as particularly pressing compared to other community problems. The salience of gender injustice may also have been lower due to the all-girls setting.

## Tensions between asset-based, open-ended opportunities for applying AI knowledge

In an effort to center learners' interests, values, and prior knowledge, we also aimed to understand the benefits and challenges of how AI literacy could be supported using an open-ended and asset-based approach. After learners came up with ideas of their own for AI tech in group and individual final projects, we asked them to do the same analysis activity that they did with existing digital assistant technology. When learners considered their own ideas, they were particularly excited about the possibilities for AI to solve problems in their communities. Many AI systems learners discussed, ideated, and analyzed were embodied, including robots in the Community AI Robot activity and semi-embodied digital assistants. 9 of 16 learners chose to brainstorm AI technology in the form of a robot for their individual final project, perhaps influenced by Activity 2's focus on robots.

In contrast to recent prior work from Druga et al. (2021), which found that children 7-11 years old prioritize having fun with AI technology over ruminating on technological drawbacks, we saw that learners who participated in our workshop could easily pinpoint harms potentially arising from existing digital assistant technology. Yet, they were less likely to come up with possible harms with their own future tech ideas compared to when analyzing existing digital assistant tech. In both the Community AI Robot activity and final projects, we dedicated prompts and slides to brainstorming potential harms. In the group activities, we asked learners to identify as many benefits and harms as possible. 5 harms on average were identified for existing digital assistant technology, while only two harms on average were identified for future robot AI technology. Three explanations for this may be that it was harder to think of harms with a futuristic technology compared to one that already exists, it is difficult to consider how one's own idea may have negative impacts, or this may relate to the finding that learners were less likely to mention humans in general as the responsible for unfair interactions with AI.

However, three learners in the follow-up interviews reasoned that AI has at least the collective knowledge of its developers. We infer that learners may connect that AI may hold biases that human techmakers have.

While learners chose problems that mattered to them, we also saw that some of the AI systems may not have lent themselves to straightforward application of AI and training data knowledge to critically analyze bias, in comparison to algorithm-focused disembodied systems (e.g., recommendations algorithm). For example, our learners who chose to discuss robots mostly came up with harms that were mistakes and/or tangible (e.g., a self defense instructor robot mistakenly attacking a person). Due to the focus on physical harms and embodied AI, it was sometimes difficult to then apply training data concepts as solutions, as we describe more next.

## Discussion

In this work, we approached AI education with a lightweight workshop approach including interactive group activities, in which learners analyzed existing and future AI technologies. Compared to other programs, our workshop content and activities included group discussions and ideation, with less use of interactive tools, as a means to understand opinions, perceptions, and knowledge gaps. We prioritized understanding algorithmic bias and help vs. harm of AI in our content and activities over technical aspects of machine learning and training data. We observed that while the lightweight workshop and activities did not offer an in-depth AI and ethics education, it appeared to `plant a seed,' with some learners we interviewed having walked away with a shift in their conceptions and gained more complex ideas about AI and related topics, as observed in follow-up interviews and final projects after less than 3 hours. This lightweight approach may be more accessible to learners who could otherwise not attend a longer, all-inclusive AI education program. Further, learners were able to understand important, overarching patterns and concepts related to algorithmic bias.

Limitations include the limited generalizability of our results by our small, self-selected sample. This may have had an impact on the types of challenges and solutions that learners focused on throughout the workshop sessions. Finally, as is common in design studies, we noticed that researcher examples (e.g., an example about accent recognition by digital assistants led learners to discuss language barriers) for specific prompts and the prompts themselves (e.g., 'how might this AI do harm') may have shifted learner responses.

Design Implications: Grounded in our data and findings, we recommend three overarching aspects to keep in mind when designing AI education opportunities that center historically excluded learners. First, we suggest ***emphasizing unanticipated bias as in all technologies, and specifically as it relates to data-driven ways AI could harm***. From our study, we saw that many learners paralleled fairness with kindness and unfairness with intentional evil, so learning content may need to directly address this mental model to be more flexible. Second, ***give training data more time and exploration***. If a learning goal is for participants to have a more in-depth, technical understanding of training data concepts, more opportunities to experiment and build a stronger mental model is needed, i.e., our lightweight approach is not as effective for particularly complex or technical learning goals in AI literacy. This is inline with prior work that presents the benefits of interactive tools (Hiltron et al., 2019). However, a comprehensive understanding of training data may not be needed for learners to understand high level patterns and concepts, since we found that learners demonstrated understanding and engaged in discussions in our lightweight approach. Third, ***opportunities to apply training data and AI bias in open-ended activities***. Since some of the topics that learners choose in an open-ended prompt may not lend themselves as easy AI and bias learning examples, we encourage instructional designers to create opportunities for more straightforward application. This may include a more specific, scaffolded prompt for learners to think of community challenges they would like to solve with AI, or breaking down their future AI ideas into considering different sensors as suggested by Touretzky et al. (2019) and considering bias from data for the sensors. Future work may also incorporate different types of bias in AI, e.g., from developers and algorithm design. We believe that these findings will help support the creation of inclusive, accessible, learner-centered opportunities for middle school girls to increase their AI literacy, engage in conversations around AI, and become empowered techmakers.

**Acknowledgments**

We thank the learners who took part in this study and our collaborators from the University of Pittsburgh, Arizona State University, and the University of Maryland Baltimore. This work is funded by NSF DRL-1811086.